\documentstyle[aps,epsfig,psfrag]{revtex}
\title{A microscopic model for the intrinsic Josephson tunneling in high-{$T_C$} 
superconductors}

\author{O.\ Schmitt, H.\ Endres and W.\ Hanke}
\address{Institute for Theoretical Physics, Universit\"at W\"urzburg, 97074 W\"urzburg 
FRG}
\author{R.\ Kleiner and P.\ M\"uller}
\address{Physikalisches Institut III, Universit\"at Erlangen-N\"urnberg, 91058 Erlangen
 FRG}

\begin{document}

\maketitle

\begin{abstract}
A quantitative analysis of a microscopic model for the intrinsic Josephson effect in 
high-temperature superconductors based on interlayer tunneling is presented. The 
pairing correlations in the $CuO_2$-planes are modelled by a 2-D  Hubbard-model with 
attractive interaction, a model which accounts well for some of the observed features 
such as the short planar coherence length. The stack of Hubbard planes is arranged on 
a torus, which is threaded by a magnetic flux. The current perpendicular to the 
planes is
calculated as a function of applied flux (i.\ e.\ the phase), and -- after careful 
elimination of finite-size effects due to single-particle tunneling -- found to display
a sinusoidal field dependence in accordance with interlayer Josephson tunneling.
Studies of the temperature dependence of the supercurrent reveal at best a mild 
elevation of the Josephson transition temperature compared to the planar 
Kosterlitz-Thouless temperature. These and other results on the dependence of the 
model parameters are compared with a standard BCS evaluation.
\end{abstract}

\centerline{PACS numbers: 74.20.-z,74.20.Mn,71.10.Fd}

\section{Introduction}
\label{intro}
The high-$T_C$ superconductors (HTSC) reveal a number of unusual properties.
One unique example, important both from a fundamental and from an applied 
point of view, is the intrinsic, ``microscopic'' Josephson effect:
a single crystal like $Bi_2Sr_2CaCu_2O_8$ (BSCCO) consists of natural stacks 
of thousands of Josephson junctions, with the $CuO_2$ layers acting as 
superconducting electrodes and the $Bi_2O_3$ layers as insulating 
(``intrinsic'') barriers, closely packed on a microscopic length 
scale ($d \approx \ 1 {\rm \AA}$). This intrinsic Josephson effect was first discovered
for BSCCO- \cite{kle92,kle94} and later confirmed in $YBaCuO$- \cite{lin95}
materials. From the point of view of the microscopic theory of the HTSC,
 this is an important effect for the c-axis transport and possibly the pairing
 theory. 
It is also an interesting and promising effect from an applied, technological 
point of view, using the microscopically packed Josephson junctions as small 
scale, tunable high-frequency oscillators \cite{mul92}.

Strong, Clarke and Anderson argued in a series of papers 
\cite{cla94,cha95,and94} that the unusual c-axis resistivity data obtained for 
cuprate superconductors are the result of the non-Fermi-liquid nature of the 
in-plane ($CuO_2$) ground state of these materials. This non-Fermi-liquid 
nature is taken as responsible for disrupting the interplane hopping of 
electrons so strongly that single electrons effectively do not hop 
between the planes, giving rise to anomalous c-axis transport properties 
\cite{cla94} and to the anomalously large superconducting transition 
temperatures for these materials \cite{cha95}. In this theory, both of these 
effects depend on the absence of normal interplane hopping. 
It is therefore natural to study the microscopic 
Josephson coupling between $CuO_2$-planes, which exhibits this characteristic.

A theoretical description of the intrinsic Josephson current must explain 
both the Kosterlitz-Thouless-type of superconductivity in the planes 
\cite{mor91,ran92,fak94} and the effective coupling of these planes 
on a microscopic basis. In previous theoretical studies of the intrinsic 
Josephson effect, the supercurrent was not investigated on such a rigorous 
microscopic basis.
Rather, it was studied using macroscopic electrodynamics employing, for example,
empirically derived coupled sine-Gordon equations for stacked 
Josephson junctions \cite{kli95,kln95}.

In this paper, we present  
numerical evidence based on Quantum-Monte-Carlo (QMC) simulations for the existence 
of the supercurrent and the 
intrinsic Josephson effect on a purely microscopic basis. The simulated model 
consists of a stack of coupled Hubbard planes for which, in analogy to the 
experimental situation, each individual plane is modelled by a two-dimensional 
(2-D) short 
coherence-length superconductor \cite{mor91,ran92,ran95}, i.e. the 2-D 
attractive Hubbard model. 

A simplified mean-field (BCS) description of this model 
similar to an earlier version by Tanaka  \cite{tan94} is discussed 
in Sec. \ref{model}.  
This BCS description is similar in spirit to the conventional Josephson 
description in that it introduces, in analogy to the 
standard treatment, the phase dependence of the BCS order parameter (which is 
constant in a plane) 
perpendicular to the planes. Despite its limitations, this simple mean-field solution 
illustrates some of the physics: in particular, the dependence of the supercurrent 
on the microscopic parameters, i.\ e.\ the inter-plane hopping  and the on-site 
interaction $U$. However, it cannot account for the  
Kosterlitz-Thouless nature of the superconductivity in the planes 
\cite{mor91,ran92,ran95} and the corresponding power-law pairing correlations.

Sec. \ref{qmc} summarizes the QMC evidence for the intrinsic Josephson effect, 
including a finite-size study with respect to the number of coupled planes
(up to 8). The influence of model parameters such as 
the temperature on the Josephson current is extracted. Of particular interest 
in view of the above-mentioned ``Josephson-pairing'' theory 
\cite{cla94,cha95,and94} is that we find at best a mild elevation 
(for intermediate interaction $U=-4 t_p$) of the Josephson 
transition temperature compared to the 2-D Kosterlitz-Thouless temperature.
Following an argument of Ferrel \cite{fer88}, this can be understood
within a weak-coupling (small $U/t_p$) BCS framework: in the normal
state, single-electron tunneling through the barrier between the two
planes lowers the energy of the system. A part of this self-energy is 
lost when the planes become superconducting because of the gap in the
quasiparticle spectrum. The lost self-energy is, however, restored by
the tunneling of Cooper pairs and the resulting Josephson coupling
energy. In the Strong, Clarke and Anderson scenario the self-energy
loss due to single-particle tunneling is assumed to be suppressed and
only the energy lowering due to Copper pair tunneling is effective.
Finally Sec.\ \ref{summ} summarizes the results.

\section{Description of the model and mean-field evaluation}
\label{model}

The superconducting $CuO_2$ planes in x-y-direction (Fig.\  \ref{torus}) 
are simulated by a 
2-D Hubbard model with attractive interaction, which is subjected to 
periodic boundary conditions:
\begin{eqnarray}
  H_p&=& -t_{p}\sum_{<i,j>_\parallel,\sigma} (c^\dagger_{i,\sigma} c_{j,\sigma}^{ }
         + {\rm{h.c.}}) \nonumber\\
         && - U\sum_i c^\dagger_{i,\uparrow}   c_{i,\uparrow}^{ }
               c^\dagger_{i,\downarrow} c_{i,\downarrow}^{ }
      -\mu\sum_{i,\sigma} c^\dagger_{i,\sigma} c_{i,\sigma}^{ }.
\label{hub1}
\end{eqnarray} 
The summation $<i,j>_\parallel$ is taken over nearest neighbors in 
the planes and $t_p$, $U>0$ and  $\mu$ denote the transfer energy in the plane, the 
attractive interaction and chemical potential, respectively. The planes are 
coupled by a hopping term in the perpendicular (z-) direction:
\begin{eqnarray}
 H_\perp&=& -t_z \sum_{<i,j>_\perp,\sigma} c^\dagger_{i,\sigma} c_{j,\sigma}^{ }
          + {\rm{h.c.}} 
\label{hp}
\end{eqnarray}
In the z-direction twisted boundary conditions are used. Fig.\ 
\ref{torus} displays the geometry used in the QMC simulations  in which the 
Hubbard planes are stacked into a torus and threaded by a magnetic field. The magnetic 
field $\vec B$ in $x$-direction is confined to the center of the circle.
Since the wave function must be single valued, a particle going once 
around the flux line acquires a phase $\exp \left(\frac{2 \pi i \Phi}
{\Phi_0} \right)$, where $\Phi_0=\frac{h}{e}$ is the elementary flux 
quantum \cite{ryd89} and $\Phi$ is the threaded flux. Thus, the fermionic 
operators are subject to the boundary conditions:
\begin{eqnarray}
 c_{\vec i + N_{z} \vec e_z,\sigma} &=& \exp \left( \frac{2 \pi i \Phi}{\Phi_0}
 \right) c_{\vec i,\sigma} \label{bound1} \\
 c_{\vec i + N_{x} \vec e_{x(y)},\sigma} &=&  c_{\vec i,\sigma}. \label{bound2} 
 \end{eqnarray}
 Here $\vec e_{x,y,z}$ are the lattice vectors of unit length and $N_z$ and 
 $N_x$ are the linear lengths in the $z$- and the $x$-directions, respectively. 
 Through a canonical transformation, the total Hamiltonian $H=H_p+H_\perp$ may be 
 written as
  \begin{eqnarray}
 H&=& -t_p\sum_{<i,j>_\parallel,\sigma} (c^\dagger_{i,\sigma} c_{j\sigma}^{ }
         + {\rm{h.c.}}) \nonumber\\
   && -t_z \sum_{<i,j>_\perp,\sigma} \left(c^\dagger_{i,\sigma} c_{j\sigma}^{ }
         \exp{\left(\frac{2 \pi i}{N_z} \frac{\Phi}{\Phi_{0}}\right)} + 
{\rm{h.c.}}\right) \nonumber\\
   && -U\sum_i c^\dagger_{i,\uparrow}   c_{i,\uparrow}^{ }
               c^\dagger_{i,\downarrow} c_{i,\downarrow}^{ }
      -\mu\sum_{i,\sigma} c^\dagger_{i,\sigma} c_{i,\sigma}^{ }.
\label{3dhubc}
\end{eqnarray}  
The fermionic operators now satisfy periodic boundary conditions in both 
the $x$- and $z$- directions. This model can be interpreted as a torus (shown 
in Fig.\ \ref{torus}) built up from stacks of coupled Hubbard $x$-$y$-planes
and threaded by a $\vec B$-field in the $x$-direction. As 
shown in Sec.\ \ref{qmc}, this geometrical arrangement is directly accessible 
to QMC simulations.

The influence of the $\vec B$-field on the Josephson dc-current is 
described by the gauge-invariant form of the Josephson equation 
\cite{and63,mah90}:
\begin{eqnarray}
j&=&\sin \left(\theta_2 - \theta_1 - \frac{2 e}{\hbar} \int(\vec A \cdot 
{\rm{d}}\vec l )\right) .
\label{jphi1}
\end{eqnarray}
Here $\theta_2 - \theta_1$ denotes the phase difference at zero magnetic 
field of the two planes considered and $\vec A$ gives the vector potential 
induced by $\vec B$.

In our microscopic model the bare interplane hopping $t_z$ in Eq.\ \ref{hp}
is real, (i.\ e.\ the bare tunneling matrix elements). In this case 
$\theta_2-\theta_1=0$, and the sinusoidal dependence of the tunneling 
current stems entirely from the line integral $\int \vec A \cdot {\rm d}\vec l
=\Phi$, i.\ e.\ the magnetic flux:
\begin{eqnarray}
j&=&\sin \left(4 \pi \frac{\Phi}{\Phi_0}\right).
\label{jphi} 
\end{eqnarray}
In order to verify this intrinsic Josephson behavior, the tunneling current 
perpendicular to the planes (in $z$-direction, Fig.\ \ref{torus}) can 
be calculated, using the relation \cite{wag91}
\begin{eqnarray}
\langle \vec j \rangle = \left \langle \frac{\partial H}{\partial \vec 
A } \right \rangle .
\end{eqnarray}
This current should display a sinusoidal 
dependence on the penetrating flux $\frac{\Phi}{\Phi_0}$, with period $\frac{1}{2}$.

Let us illustrate the idea by considering first a simplified BCS treatment 
\cite{tan94}, which is based on a Hartree evaluation of the free energy $F$ and the 
current 
\begin{equation}
j = \left\langle \frac{\partial H}{\partial A} \right\rangle \propto 
\frac{\partial F}{\partial (\Phi/N_z)} .
\label{strom}
\end{equation}
Here, the interaction part 
\begin{equation}
H_{\rm int}= -U\sum_i c^\dagger_{i,\uparrow}   c_{i,\uparrow}^{ }
               c^\dagger_{i,\downarrow} c_{i,\downarrow}^{ }
              \nonumber
\end{equation}
is simplified to the usual BCS form 
\begin{eqnarray}
H_{\rm int} \approx -\sum_i\left(c_{i\uparrow}^\dagger c_{i 
\downarrow}^\dagger \Delta + {\rm h.c.} - \frac{\vert \Delta \vert 
^2}{\vert U \vert}\right),
\label{ww}
\end{eqnarray}
where $\Delta= \vert U \vert \langle c_{i \uparrow} c_{i \downarrow} \rangle $ 
denotes the pair potential. For an attractive on-site potential, the 
Cooper pairs have simple s-wave symmetry \cite{ran92}. The usual phase dependence is
introduced by writing the pair potential $\Delta_m=\Delta  e^{im\phi}$ for the $m$-th 
plane. 

Fourier-transforming in all spatial directions, this BCS Hamiltonian then can be written:
\begin{eqnarray}
 H = \sum_{k\sigma}{\epsilon(k) c^\dagger_{k\sigma} c_{k\sigma}} 
 -\sum_k{\Delta(k) [c^\dagger_{k \uparrow} c^\dagger_{-k \downarrow} +
 c_{-k \downarrow} c_{k \uparrow}]} 
+ \sum_k \frac{\Delta^2(k)}{\vert U \vert},
\label{hambi} 
\end{eqnarray} 
where $k=(k_x,k_y,k_z)$ is the 3-D wavevector and 
\begin{eqnarray}
\epsilon(k)=-2t_p[\cos(k_x a)+\cos(k_y a)] - 2t_z 
\cos\left(k_z a+\left(\frac{2 \pi}{N_z} \frac{\Phi}{\Phi_{0}}\right)\right)    
\label{eps}
\end{eqnarray}
gives the kinetic energy of the electrons. This Hamiltonian, which is 
bilinear in its fermionic operators, can be diagonalized using a standard 
Bogoliubov transformation:
\begin{eqnarray}
 H = \sum_k{E_k(\alpha_{k \uparrow}^\dagger \alpha_{k \uparrow} +
 \alpha_{k \downarrow}^\dagger \alpha_{k
\downarrow})} - \sum_k{[\tilde E_k - \tilde \epsilon_k]} + 
\sum_k{\frac{\Delta^2}{U}}, 
\end{eqnarray}
where
$ \tilde \epsilon_k = \frac{1}{2}[\epsilon_{+}(k)  + \epsilon_{-}(k)], $ 
$ \tilde E_k = \sqrt{\tilde \epsilon_k^{2} + \Delta^2} $  and
$ E_k = \tilde E_k + \frac{1}{2}[\epsilon_{+}(k) - \epsilon_{-}(k)] $
is the BCS quasiparticle-energy, and
\begin{eqnarray}
 \epsilon_{+}(k) &=& 2t_p(\cos(k_x a) + \cos(k_y a)) +
 2 t_z \cos\left(k_z a + \left(\frac{2 \pi}{N_z} \frac{\Phi}{\Phi_{0}}\right)\right) -
 \mu \nonumber \\
 \epsilon_{-}(k) &=& 2t_p(\cos(k_x a) + \cos(k_y a)) +
 2 t_z \cos\left(k_z a - \left(\frac{2 \pi}{N_z} \frac{\Phi}{\Phi_{0}}\right)\right) -
 \mu .
\end{eqnarray}

The free energy can be obtained analytically from $H$, yielding:
\begin{eqnarray}
F = \frac{N \Delta^2}{U} - 
\sum_k{(\tilde E_k - \tilde \epsilon_k)} - 2 
k_{\rm{B}} T \sum_k{\log(1+e^{-\beta  E_k })}. 
\end{eqnarray}
The tunneling current $j(\Phi)$ can be obtained by numerically 
differentiating $F(\Phi)$. The parameter $\Delta$ is extracted by 
minimizing the free energy, resulting in the BCS-type gap equation:
\begin{eqnarray}
  \Delta = \frac{U}{N} \sum_k{\frac{\Delta}{2 \tilde E_k}
(1 - f(E_k) - f(E_{-k}))}. 
\end{eqnarray}

A typical result of this simplified analysis is shown in Fig.\ 
\ref{jphi1b}, which illustrates the dependence of the mean-field 
Josephson current on the microscopic parameters, $t_z$ and $U$ 
(here and in what follows all energies are measured in units of $t_p$). 
In Fig.\  \ref{jphi1b} the current $j(\Phi/\Phi_0)$ is plotted against 
the magnetic flux $\Phi/\Phi_0$ for $t_z=0.1 t_p$ at half filling and 
for various values of 
$U$. The current shows the expected sinusoidal $\Phi$-dependence with 
period $\frac{1}{2}$. The maximum current decreases with increasing on-site 
interaction, because the effective mass of a Cooper pair and thus its 
localization tendency in a plane scales roughly with $U$ \cite{tan94}. 
If the interaction drops below a  critical value, the 
supercurrent vanishes above a critical  value of $\Phi$. This behavior can be 
understood by looking at the energy of the Bogoliubov quasi-particles, 
which can be written as
\begin{eqnarray}
  E_k =\sqrt{\tilde \epsilon_k^2 + \Delta^2(k)} 
  - 2 t_z \sin(k_z a) \sin \left( \frac{2 \pi}{N_z} \frac{\Phi}{\Phi_0} \right) 
\label{ek3} .   
\end{eqnarray}
 If the pair potential $\Delta$ is smaller than the hopping 
 energy $t_z$, the energy $E_k$ becomes negative for large $\Phi$ and, 
 therefore, a quasi-particle is excited, which corresponds to the 
 destruction of a cooper pair.
 One should notice that for $\Phi=0$, i.\ e.\ without an applied $\vec B$-field,
 $E_k$ is positive for all finite 
 values of $\Delta$, and BCS
 superconductivity is stable for all finite values of $U$. 
 Similarly, one can extract the influence of the filling $\langle n\rangle$ 
 on the supercurrent $j(\Phi)$. As expected, $j(\Phi)$ scales with the number of 
 electrons in the system.

 What goes wrong in this mean-field approach? 
 In 2-D systems, fluctuations drive the critical 
 temperature $T_C$ for the onset of  off-diagonal long-range order (ODLRO) 
 down to zero \cite{sch89}. There exists, 
 however, a Kosterlitz-Thouless transition at a temperature $T_{KT}$, below which 
 there is quasi-long-range-order (power-law pair correlations) \cite{kos73}. 
 Due to experimental evidence for 
 intra- and inter-unit-cell Josephson junctions in $YBaCuO$ single crystals 
 \cite{lin95}, it is generally believed that the high-$T_C$ superconductivity
 is intrinsically 2-D in $CuO_2$ bilayers, coupled together by Josephson currents
 along the c-axis (our $z$-axis) direction. There is also clear experimental 
 evidence from the short 
 coherence length (a few lattice parameters) that the 2-D superconductivity is
 different from the conventional BCS one: the typical radius ( $\approx$ coherence 
 length) of a Cooper-pair is comparable to the distance between pairs, and in this
 sense the HTSC are in an intermediate regime between the usual BCS weak-coupling 
 regime and a regime in which pre-existing pairs Bose condense \cite{tok91,dre90}.
 QMC simulations \cite{mor91,ran92} have correctly established the 
 short-coherence length Kosterlitz-Thouless type of superconductivity for the 2-D
 attractive Hubbard model. It is therefore natural, to also apply this numerically 
 rigorous QMC 
 procedure to search for numerical evidence for Josephson tunneling in a 
 stack of 
 coupled $CuO_2$ planes.

 \section{QMC results}
 \label{qmc}

 The Josephson current is characterized by the sinusoidal dependence 
 $j(\Phi)=J_0 \sin\left(4 \pi \frac{\Phi}{\Phi_0}\right) $. In our 
 actual QMC simulation (see below), we detect also current contributions of the 
 form:
 \begin{eqnarray}
j_s(\Phi)= J_s\sin \left( 2 \pi \frac{\Phi}{\Phi_0} \right),
\nonumber
\end{eqnarray} 
 which are due to single-particle tunneling. This is a ``finite-size'' 
 effect, with $ \lim_{N_z \to \infty} J_s = 0 $,
which has to be separated from the physical effect we are after.

The current
\begin{eqnarray}
 \langle \vec j \rangle &=& \left\langle \frac{\partial H(\vec A)}
{\partial \vec A}\right\rangle \label{jphys}
\end{eqnarray}
perpendicular to the planes can be calculated using a finite-temperature QMC 
technique \cite{loh90} for the geometry of Fig.\ \ref{torus}.
The expectation value in equation (\ref{jphys}) can directly be expressed in 
Green's functions $G_{ij}$, which are accessible to standard QMC routines, i.e.
\begin{eqnarray}
  j&=&- t_z \sum_{<i,j>_z\sigma} \left[ c_{i\sigma}^\dagger c_{j\sigma} 2 \pi i 
 \frac{N_z}{\Phi_0 L} \exp\left(-2 \pi   i \frac{\Phi}{\Phi_0} \frac{1}{N_z}
\right) -\rm{h.c.} \right] \nonumber \\
 &=&-i t_z \frac{2\pi}{a \Phi_0} \left[\sum_{<i,j>_z,\sigma}
c_{i\sigma}^\dagger
c_{j\sigma} \exp(-2 \pi  i \frac{\Phi}{\Phi_0} \frac{1}{N_{z}}) -\rm{h.c.} \right] ,
\end{eqnarray}
where $\Phi= A_z L$ with $L=N_{z} a$ and $a$ is the lattice parameter. Using
\[ \langle c_{i\sigma}^\dagger c_{j\sigma} \rangle= 
\delta_{ij} - \langle c_{j\sigma} c_{i\sigma}^\dagger 
\rangle = \delta_{ij}-G_{ji}^{\sigma}, \]
the current becomes:
\begin{eqnarray}
\langle j \rangle &=&- i t_z \frac{2 \pi}{a \Phi_0} 
\sum_{<i,j>_z\sigma} \left[ G_{ij}^{\sigma} \exp \left(2 \pi i 
\frac{\Phi}{\Phi_0} \frac{1}{N_z} \right) - G_{ji}^{\sigma} 
\exp \left(-2 \pi i \frac{\Phi}{\Phi_0}\frac{1}{N_z} \right) \right]. 
\end{eqnarray}

In the first simulation we used a $4\times 4 \times 4$ lattice with periodic boundary 
conditions in all three spatial directions. Because of the limited system size, the
results display considerable finite-size effects. In order to separate these 
finite-size effects 
from the physical effect we are after, we have calculated $j(\Phi)$ in the 
non-interacting ($U=0$ in eq.\  (\ref{3dhubc})) tight-binding model for two
different sizes  (Fig.\ \ref{u01}). The results show, that the finite-size effects 
are also sinusoidal, but with period 1, and, indeed (see the argumentation below), 
they are related to single-
particle tunneling. Thus, if a Josephson current appears in the system, $j(\Phi)$ 
should behave like a superposition of two sinusoidal curves with period $\frac{1}{2}$ 
and 1. 
It should, therefore, be of the following additive form
 \begin{eqnarray}
j(\Phi)=J_j \sin\left(4 \pi \frac{\Phi}{\Phi_0}\right)+ 
J_s \sin\left(2 \pi \frac{\Phi}{\Phi_0}\right),
\label{jsoll}
\end{eqnarray}  
with the first term referring to the Josephson supercurrent and the second to the 
single-particle finite-size effects. The parameters $J_j$ and $J_s$ are the 
corresponding amplitudes. 

The Hubbard model parameters used in the following discussion are: $U=-4 t_p$, $\langle n \rangle=0.75$, 
$t_z=0.1 t_p$, for which the 2-D attractive Hubbard planes are known to show 
Kosterlitz-Thouless type of superconductivity for $\beta \cong 8$ -- 10 
\cite{fak94,mor91}.

As seen from Fig. \ref{u41}, which contains the QMC results (diamonds plus 
errorbars), the curve (\ref{jsoll}) fits the calculated points very well. Thus, we 
have a first evidence that our model indeed describes a Josephson junction. 

By increasing the number of planes from 4 to 8, the amplitude, $J_s$, of the 
single-particle finite-size effects can be reduced (Fig.\ \ref{u42}), and the sinusoidal 
curve with period $\frac{1}{2}$ can 
be seen more clearly. The interpretation of 
$j_s(\Phi)=J_s \sin\left( 2 \pi \frac{\Phi}{\Phi_0}\right)$ as a finite-size 
contribution, that is related to the single-particle tunneling, is based on its scaling 
behavior as a function of the number of $N_z$ of planes. 
The reason for this finite-size behavior of $j_s$ can simply be understood: 
Enhancing the number of coupled planes, $N_z$, the phase of the single-particle 
hoppings in eq.\ (\ref{3dhubc}) is more effectively averaged out to zero than for smaller 
$N_z$, and $J_s$ should scale to zero. 

On the other hand, $J_j$, the Josephson amplitude, should scale with the number of planes
$N_z$, a relation, which is clearly obeyed by our QMC data in Figs.\ \ref{u41} and 
\ref{u42}. In Fig.\ \ref{u41} and \ref{u42},  $\frac{1}{N_z} j(\Phi)$ 
is plotted against $\frac{\Phi}{\Phi_0}$ for  
 $4 \times 4 \times 4$ and $4 \times 4 \times 8$ systems, respectively. 
The lines indicate the interpolated curve. As expected, the magnitudes of 
$\frac{1}{N_z}J_j$ in 
Figs.\ \ref{u41} and \ref{u42} are equal ($J_j=0.0079$ is used in both figures). 

The 2-D attractive Hubbard model for $U=-4$ and $\langle n \rangle=0.75$ 
shows a
Kosterlitz-Thouless transition at $\beta_c t_p \approx 10$ from the superconducting 
to a normal phase \cite{fak94,ran92}.
This transition should be reflected in coupled Hubbard planes, 
and the Josephson current should break down at some $\beta=\beta_c$. 
The temperature dependence of $j$ allows an estimation of 
$\beta_c$. In Figs.\ \ref{u45}, \ref{u46} and \ref{u47}, $j(\Phi)$ is plotted 
for $\beta=2,6$ and 7 respectively. Fig.\ \ref{u41} gives the 
corresponding plot for $\beta=8$. For $\beta=2$, $j(\Phi)$ is 
sinusoidal with period $1$, i.\ e.\  no Josephson current appears, and the  
system is not superconducting. For $\beta=6$ and $\beta=7$, $J_j$ 
becomes finite, but is still small compared with the value of $J_j$ for 
$\beta=8$. 
The $J$ values for $\beta=10$ (not plotted) have been found to be essentially 
unchanged compared to $\beta=8$. This growing current  
(with saturation around $\beta \approx 8$) is due to interlayer tunneling of local 
pairing of electrons. 
For temperatures $\beta \le 8$, exponential not power-law pairing correlations are present
in the 2-D Hubbard model. Accordingly, preliminary results indicate 
that for $\beta<8$ the current does not scale with the number of 
sites in a plane.
A rough estimate for the Josephson transition temperature is, therefore, $\beta 
\cong 8$, at best a mild elevation compared to the single plane.
Within BCS Ferrel showed that the shift in ground-state energy due to
tunneling is the same in the normal and superconducting stats
\cite{fer88}. The physical significance of this result is that the
ground-state energy of the junctions in its superconducting state is
neither favored nor disfavored, relative to the normal-state energy by
the presence of the tunneling barrier. Therefore, in this BCS limit,
we do not expect any elevation in the transition temperature.

\section{Conclusions}
\label{summ}

In this paper, we have presented a model for intrinsic Josephson couplings  
on a microscopic length-scale. It is based on coupled  Hubbard planes with 
attractive interactions, which are threaded 
by a local magnetic field. The field dependence of the current 
$j(\Phi)$ perpendicular to the planes is calculated using numerically rigorous
Quantum-Monte-Carlo simulations. To extract some of the physical ideas, in particular
the dependence on the microscopic parameters $U/t_p$ and $t_z/t_p$, we first reviewed
the mean-field BCS approximation. The maximum current $j_{\rm max}$ increases with 
decreasing on-site 
interaction $U$ and increasing hopping energy $t_z$. At small 
values of $U$, for which the pair potential $\Delta$ is smaller than 
$t_z$, the superconductivity is unstable for finite values of the 
magnetic flux $\Phi$. In mean-field, the Hubbard model reduces to an 
effective BCS type Hamiltonian, which is not capable of correctly describing salient
features of the 2-D superconductivity (short coherence length, Kosterlitz-Thouless 
transition) in
the $CuO_2$ planes of the HTSC's. 

To overcome these shortcomings and to present a 
rigorous proof of the intrinsic Josephson couplings, we have extracted $j(\Phi)$ 
from the,
in principle (apart from controlled statistical error \cite{loh90}), exact 
Quantum-Monte-Carlo simulations. The results show a sinusoidal behavior of 
$j(\Phi)$. A careful discussion of the finite-size scaling of the current clearly 
reveals that it is the behavior of a Josephson supercurrent. The Josephson current 
drops to zero for $\beta<6$, at $U=-4$ and a filling $\langle n \rangle =0.75$. 
Due to finite-size effects, this transition is smeared out and, therefore, 
we can only
approximatively determine the transition temperature from superconducting to the 
normal  state at $\beta \cong 8$. These facts give 
clear evidence that the attractive Hubbard model is capable of 
describing the Josephson effect in HTSC's on microscopic length scale. 
Careful studies of the parameter dependence of the maximum supercurrent are under
way. These studies should enable us to systematically extract the influence of 
material properties
on the Josephson current, and thus should be useful to optimize the intrinsic Josephson
effect for specific applications.
 
\section{ Acknowledgments}
The authors would like to thank D.\ J.\ Scalapino and T. Hagenaars for many
valueable discussions.
One of us (W.\  Hanke) gratefully acknowledges the hospitality of the Physics 
Department
of the University of California Santa Barbara, where some of this work was carried out.
This work was supported by the Bavarian program FORSUPRA.

{\normalsize
\begin{figure}
\vspace*{2.0cm}
\begin{center}
\psfrag{bfeld}{$\vec B$}
\psfrag{st}{$\vec j$}
\psfrag{zr}{$\vec z$-direction}
\psfrag{Eb}{plane in $x$-$y$-direction}
\psfrag{x}{$ $}
\psfrag{y}{$ $}
\epsfig{file=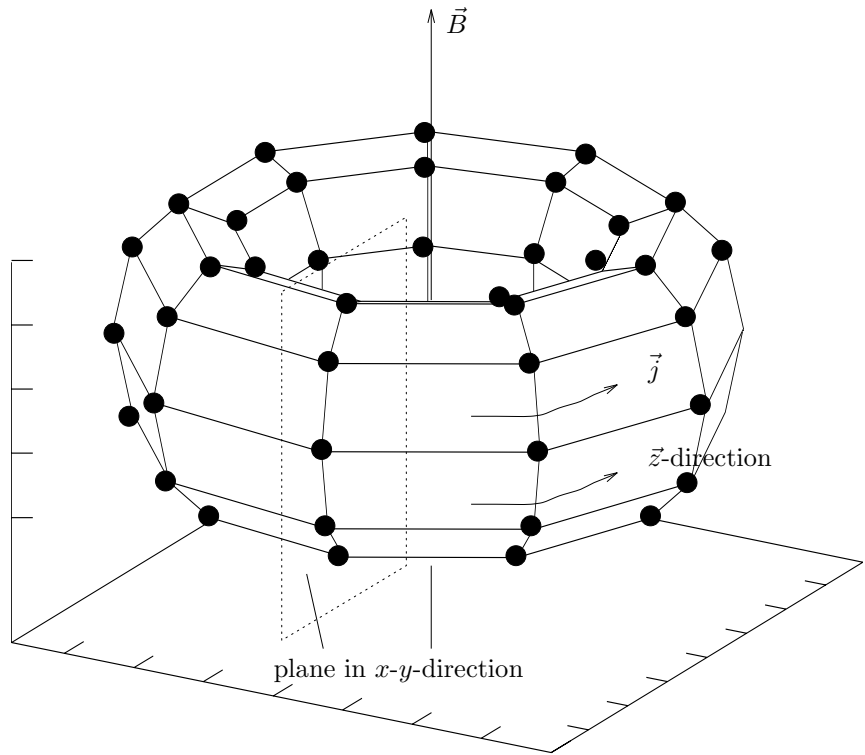,width=12cm}
\end{center}
\caption{Schematic view of the used geometry. The coupled superconducting 
planes are arranged in a torus threaded with a the local magnetic field $\vec B$.} 
\label{torus}
\end{figure}
\newpage
\begin{figure}
\begin{center}
\input{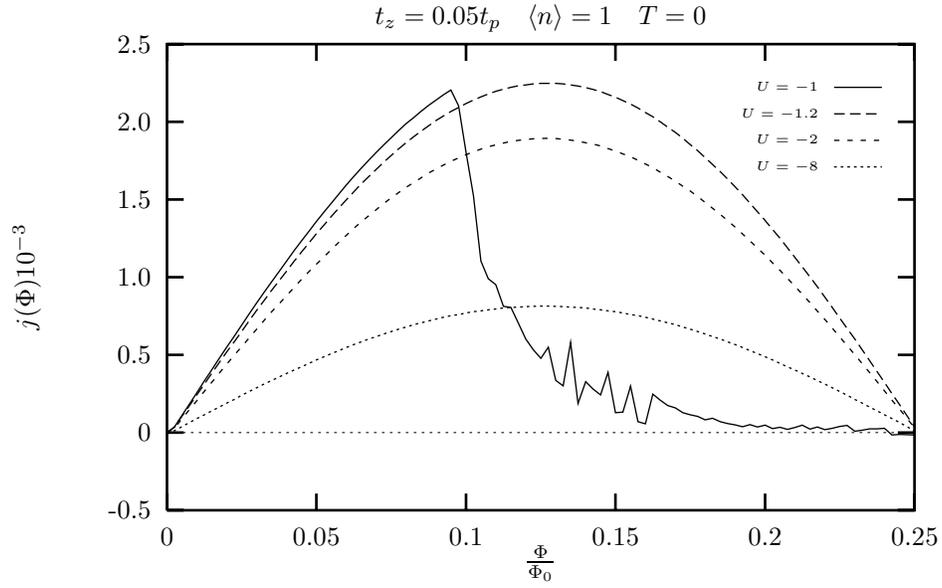}
\end{center}
\caption{Mean-Field Current perpendicular to the planes versus magnetic flux for 
$t_z=0.05 t_p$ and various values of $U$ at $\frac{1}{2}$-filling $\langle n \rangle=1$ and $T=0K$. 
The oscillations for $U=-1$ are due to the numerical differentiation.}
\label{jphi1b}
\end{figure}

\begin{figure}
\begin{center}
\input{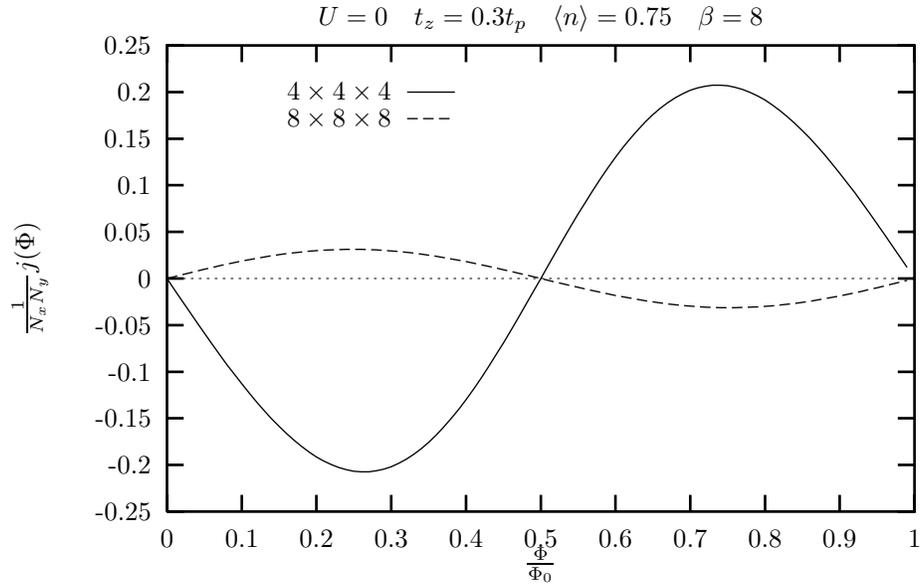}
\end{center}
\caption{Current perpendicular to the plane for the non-interacting ($U=0$) 
tight-binding model.}
\label{u01}
\end{figure}
\begin{figure}
\begin{center}
\input{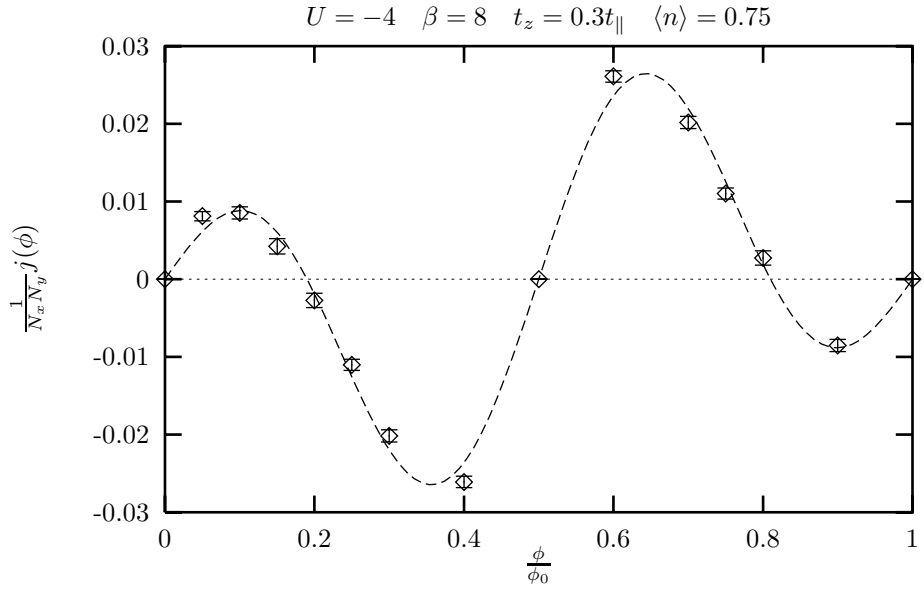}
\end{center}
\caption{Current perpendicular to the plane for the $4 \times 4$ (planar size) 
$\times 4$ (number of planes) system QMC-results: diamonds plus errorbars, 
the dotted line shows the interpolated curve $j(\Phi)$.}
\label{u41}
\end{figure}
\begin{figure}
\begin{center}
\input{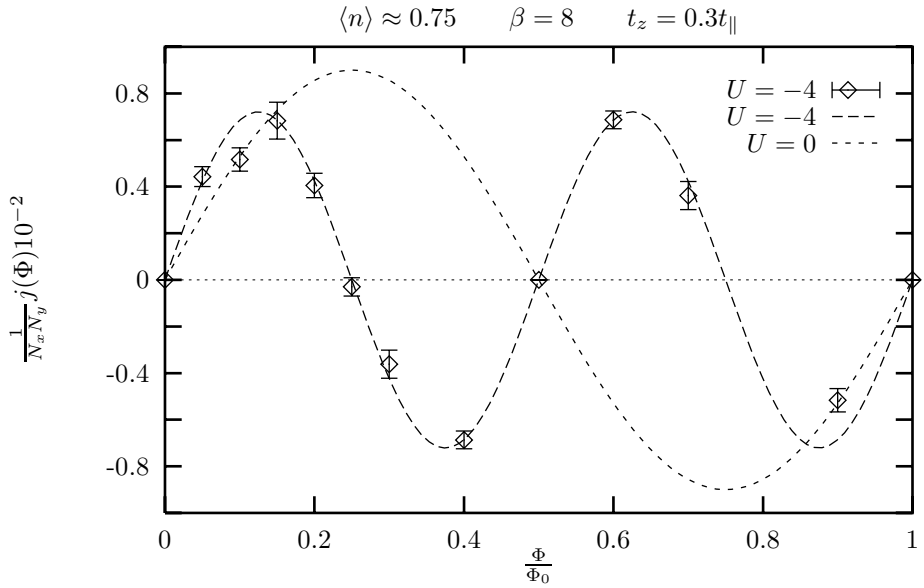}
\end{center}
\caption{Current perpendicular to the plane for the $4 \times 4 \times 8$ 
(8 coupled planes in Fig.\ \ref{torus}) system. 
The dashed line shows the $U=0$ behavior.}
\label{u42}
\end{figure}
\begin{figure}
\begin{center}
\input{u45.psl}
\end{center}
\caption{Current perpendicular to the plane for $\beta t_p=2$.}
\label{u45}
\end{figure}
\begin{figure}
\begin{center}
\input{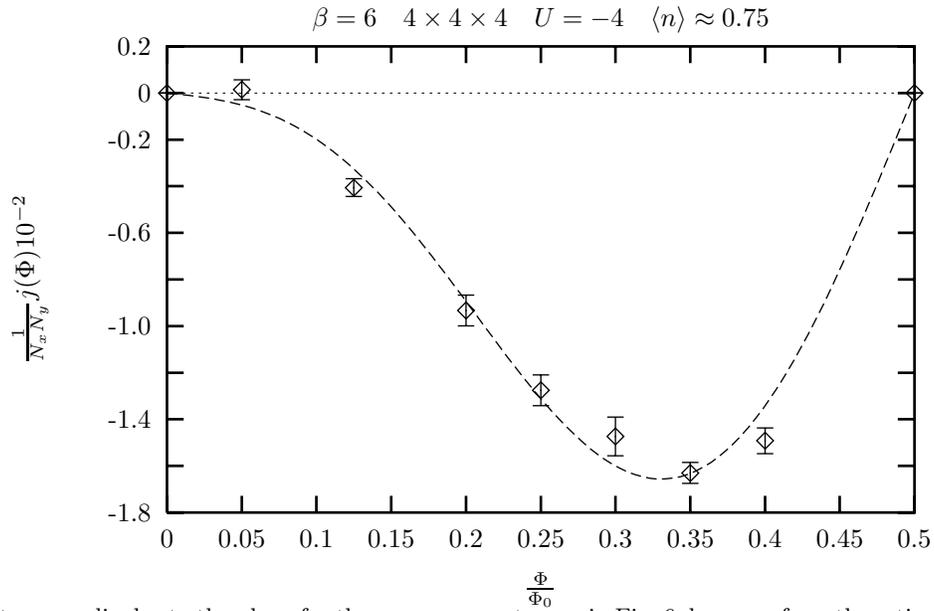}
\end{center}
\caption{Current perpendicular to the plane for the same parameters as in Fig.\ 
\ref{u45}, however for a three times lower temperature, i.\ e.\ $\beta t_p=6$.}
\label{u46}
\end{figure}
\begin{figure}
\begin{center}
\input{u47.psl}
\end{center}
\caption{Current perpendicular to the plane for $\beta t_p =7$.}
\label{u47}
\end{figure}


\begin{thebibliography}{99}
\bibitem{kle92}{R.\  Kleiner, F.\  Steinmeyer, G.\  Kunkel, P.\  M\"uller,
 Phys.\  Rev.\  Lett.\  {\bf{68}}, 2394 (1992).}
\bibitem{kle94}{R.\  Kleiner, P.\  M\"uller, Phys.\  Rev.\  B {\bf{49}},
 1327 (1994).}
\bibitem{lin95}{D. C.\ Ling, G.\ Yong, J.\ T.\ Chen and L.\ E.\ Wenger,
Phys.\  Rev.\  Lett.\  {\bf{75}}, 2011 (1995).}
\bibitem{mul92}{Hier sollte eine Anwendung stehen}
\bibitem{cla94}{D. G. Clarke, S.P. Strong and P.W. Anderson, Phys. Rev. Lett. 
{\bf{72}}, 3218 (1994).}
\bibitem{cha95}{S. Chakravarty, A. Sudbo, P.W. Anderson, S. Strong, Science
{\bf 261}, 337 (1993)}
\bibitem{and94}{P.W. Anderson, Science {\bf{256}}, 1526 (1992).}
\bibitem{mor91}{A. Moreo, D.J. Scalapino and S.R. White, Phys. 
Rev. B {\bf{45}}, 7544 (1992).}
\bibitem{ran92}{M. Randeria, N. Trivedi, A. Moreo and R.T. Scalettar, 
Phys. Rev. Lett. {\bf{69}}, 2001 (1992).}
\bibitem{fak94}{ F.F. Assaad, W. Hanke and D.J. Scalapino, Phys. Rev. B
 {\bf{50}}, 12835 (1994).}
\bibitem{kli95}{R. Kleiner et. al., Phys. Rev. B {\bf 50}, 3942 (1994).}
\bibitem{kln95}{R. Kleiner, Phys. Rev. B {\bf 50}, 6919 (1994).}
\bibitem{ran95}{N. Trivedi and M. Randeria, Phys. Rev. Lett. {\bf 75}, 312 (1995).}
\bibitem{fer88}{R. A. Ferrel, Phys. Rev. B {\bf 38}, 4984 (1988).}
\bibitem{tan94}{Y. Tanaka, Physica C  {\bf{219}}, 213 (1994).}
\bibitem{fak93}{F.F. Assaad, W. Hanke and D.J. Scalapino, Phys. Rev. Lett. 
 {\bf{71}}, 1915 (1993).}
\bibitem{ryd89}{L.H. Ryder, {\it{Quantum Field Theory}}, Cambridge 
University Press 1985.}
\bibitem{and63}{P.W. Anderson and J.M. Rowell, Phys. Rev. Lett.
{\bf{10}}, 230 (1963).}
\bibitem{mah90}{G.D. Mahan, {\it{Many-Particle Physics}}, Plenum Press 1993.}
\bibitem{wag91}{J. Wagner, W.Hanke and D.J. Scalapino, Phys. Rev. B  
{\bf{43}}, 10517 (1991).}
\bibitem{loh90}{E.Y. Loh and J.E. Gubernatis, in {\it{Electronic Phase 
Transitions}}, published by  W. Hanke and Y. Kopaev (Elsevier, New 
York 1990).}
\bibitem{sch89}{S. Schmitt-Rink, C.H. Varma and A.E. Ruckenstein, 
Phys. Rev. Lett. {\bf 63}, 445 (1989).}
\bibitem{kos73}{J.M. Kosterlitz and D.J. Thouless, J. Phys. C.
{\bf{6}}, 1181 (1973).}
\bibitem{alv92}{J. J. Vicente Alvarez and C. A. Balseiro, Phys. Rev. B {\bf 46},
11787 (1992). }
\bibitem{tok91}{A. Tokumitu et.\ al.\ , J. Phys. Soc. Jpn. {\bf{60}}, 380 (1991).}
\bibitem{dre90}{M. Drechsler and W. Zwerger, Ann. Phys. {\bf 1}, 15 (1992).}
\end{thebibliography}
\end{document}